\documentclass[11pt]{article}

\usepackage{epsfig, amsfonts, amsmath, amssymb, amsthm, pictex, verbatim, amsthm, pdfsync, mathrsfs}
\usepackage{tensind}
\usepackage{endnotes}
\usepackage[normalem]{ulem}

\usepackage{hyperref}  
\hypersetup{colorlinks=true, citecolor=blue, urlcolor=blue,linkcolor=blue} 
\usepackage{graphicx}
\usepackage[round]{natbib}
\usepackage{amsmath}
\usepackage{cases}
\usepackage{amssymb}
\usepackage{hyphenat}
\usepackage[hypcap]{caption} 

\usepackage{tikz} 					
\usetikzlibrary[patterns]				
\usetikzlibrary[arrows]				
\usetikzlibrary{positioning}			
\usepackage[color=yellow, textwidth=2.8cm, textsize=tiny, shadow, disable]{todonotes}				
\usepackage{marginnote}

\usepackage{framed}			

\usepackage{comment}

\usepackage{booktabs} 
\usepackage{multirow}
\usepackage{tabularx}
\usepackage{fixltx2e} 
\usepackage{booktabs} 
\usepackage{multirow}

\usepackage{verbatim}  

\usepackage{caption}
\usepackage{subcaption}   

\usepackage{cite}  

\usepackage[margin=1.52in]{geometry}
\usepackage{authblk}

\usepackage{xparse}

\makeatletter
\newcommand{\customlabel}[2]{ \protected@write \@auxout {}{\string \newlabel {#1}{{#2}{\thepage}{#2}{#1}{}} } \hypertarget{#1}{}}
\makeatother

\definecolor{cblue}{RGB}{100,5,255}    
\definecolor{cred}{RGB}{255,10,10} 
\definecolor{cgreen}{RGB}{5,165,20}  
\definecolor{corange}{rgb}{1.0,0.49,0.0}  
                                                                   
\title{Dark Matter Realism}
\author[1,2,3]{Niels C.M.\ Martens\footnote{\url{nmartens@uni-bonn.de}; \href{https://orcid.org/0000-0002-2839-1387}{orcid.org/0000-0002-2839-1387}}}
\affil[1]{DFG Research Unit ``The Epistemology of the Large Hadron Collider'' (grant FOR 2063), Germany}
\affil[2]{Lichtenberg Group for History and Philosophy of Physics, University of Bonn, Germany}
\affil[3]{Institute for Theoretical Particle Physics and Cosmology, RWTH Aachen University, Germany}
\date{\small Accepted manuscript version of an open access paper published in \emph{Foundations of Physics} (2022), 52(1):16: \url{https://link.springer.com/article/10.1007/s10701-021-00524-y}\\[4mm] Winner of the New Directions in Philosophy of Cosmology Essay Award (Dec 2020) \url{www.philcosmo.uwo.ca/philosophy-of-cosmology-essay-contest-winners-2020/}}
\begin{document}

\todo{Update abstract and tell production team to change the abstract in the system. Also give them my orcid ID.}

	
\maketitle

\abstract{According to the standard model of cosmology, $\Lambda$CDM, the mass-energy budget of the current stage of the universe is not dominated by the luminous matter that we are familiar with, but instead by some form of dark matter (and dark energy). It is thus tempting to adopt scientific realism about dark matter. However, there are barely any constraints on the myriad of possible properties of this entity---it is not even certain that it is a form of matter. In light of this underdetermination I advocate caution: we should not (yet) be dark matter realists. The ``not(-yet)-realism'' that I have in mind is different from Hacking's (1989) anti-realism, in that it is semantic rather than epistemological. It also differs from the semantic anti-realism of logical empiricism, in that it is naturalistic, such that it may only be temporary and does not automatically apply to all other unobservables (or even just to all other astronomical unobservables, as with Hacking's anti-realism). The argument is illustrated with the analogy of the much longer history of the concept of a gene, as the current state of the concept of dark matter resembles in some relevant ways that of the early concept of genes.}

\vspace{3mm}

\begin{quote}
	The label ``dark matter'' encapsulates our ignorance regarding the nature of most of the matter in the universe. \hfill --Avi Loeb, 2021\footnote{\label{fn1}Scientific American, 30 May 2021, \url{https://www.scientificamerican.com/article/maybe-dark-matter-is-more-than-one-thing}. Jeffrey Newman expresses a similar view with regards to dark energy: ``Dark energy is really just a name for our ignorance'' (\url{www.pitt.edu/pittwire/features-articles/astronomers-shed-light-dark-energy-smallest-black-hole}, 8 Nov 2019), which is echoed by Michael Turner (\citeyear[p.1268]{turner2018}) who calls it a ``proxy for the real explanation''. As early as \citeyear{king1977}, Ivan King said: ``Can we really claim to know anything about the nature of the universe if we don't know the properties, or even the nature, of 90 percent of its material?'' (p.9; as quoted by Peebles \citeyear[p.278]{peebles2020}). In \citeyear{peebles2020}, Peebles comments on Einstein's cosmological constant gaining a new name, `dark energy', as ``a poor disguise for a \emph{fudge factor} that we accept because it serves to unify theory and observations so well'' (emphasis mine) (p.3). Moreover, commenting on the tight accumulation of evidence for $\Lambda$CDM emboldening most to talk about $\Lambda$CDM as that what ``really happened'', he favours a more cautious interpretation: ``The notion of reality is complicated, so a more secure statement would be that whatever happened---and we assume something did happen---left traces that closely resemble those predicted by $\Lambda$CDM'' (p.6).}
\end{quote}

\section{Introduction}

Open almost any modern textbook on cosmology or astrophysics/astronomy, and it will proclaim the standard model of cosmology, $\Lambda$CDM, according to which the mass-energy budget of the current stage of the universe is \emph{not} fully accounted for by the luminous matter that we know and love.\footnote{More precisely, the particles in the standard model of particle phsyics, excluding its neutrinos.} In fact, luminous matter is, at current times, not even dominant. It contributes a mere 5\% to the total mass-energy---the remainder being accounted for by (cold) dark matter (CDM), 27\%, and dark energy ($\Lambda$), 68\%. Given this received view (bracketing for now the minority alternative, modified gravity), one may well be forgiven for making the seemingly small further step of being a (scientific) realist about dark matter and dark energy.\footnote{See Jacquart (\citeyear{jacquartcompanion}) for a brief discussion of dark matter realism. Ruphy (\citeyear{ruphy2011}) warns against realism in the context of dark matter simulations, due to the strong idealisations involved.} However, this near-consensus on $\Lambda$CDM and some form of its dominant, titular components stands in stark contrast with the disagreement over the exact nature of these components, i.e.\ both the type of `stuff' it is supposed to be (i.e.~its ontology) and its properties (i.e.~its ideology).\footnote{This distinction between ontology and ideology stems from Quine (\citeyear{quine1951}).} Both are barely restricted by the available empirical data---for instance, viable dark matter candidates span 90 orders of magnitude (!) in mass \citep{bertonetait2018}. In light of this disagreement, it behoves one to ask what it would mean exactly to be a scientific realist about such elusive `stuff'. More pointedly, can one even be a realist about x, can one say that x exists, without knowing (almost anything at all about) what x is? How does scientific realism about these dark ingredients of the universe avoid vacuousness? In this paper we will focus on realism about dark matter,\footnote{The research programme sketched in \citep{dmeditorialb}---integrating dark matter, modified gravity, and the humanities---lists three groups of research questions: semantic/metaphysical, descriptive, and normative/epistemological/methodological questions. This paper engages with (the last question of) the first group.} but it should be noted that most arguments carry over to dark energy (cf.~fn.\ref{fn1}). It will be argued that it is too early to proclaim any non-trivial, substantive scientific realism about dark matter (DM). 

Section \ref{DM} will introduce dark matter models in more detail, and argue for the semantic thinness of the core concept of dark matter that is common to these models. This situation is strikingly similar to the early concept of genes---before they were even called by that name (Section \ref{protogenes}). These similarities help illustrate the particular aspect of scientific realism that is problematic in the specific case of dark matter, namely its semantic dimension (Section \ref{semrealism}), which makes this case study independent from that of the epistemological anti-realism that has already been suggested in the general context of astronomy and cosmology. Section \ref{discussion} further evaluates the case against present-day-realism about dark matter by comparing it with the case for contemporary \emph{anti}-realism about genes, concluding in Section \ref{conclusion} that it is too early to be semantic realists, and \emph{a fortiori} full-blown scientific realists, about dark matter.

\section{Dark matter} \label{DM}

The problem with dark matter realism is not a lack of viable suggestions, i.e.~models, concerning the nature of dark matter, but rather their abundance\footnote{Cf.\ Bertone and Tait's (\citeyear{bertonetait2018}) `no stone left unturned' guiding principle for this new era in the search for dark matter.}---or, more precisely, the thin common conceptual core of this cornucopia, this ``babel'',\footnote{\citet[p.5]{diluzio2020}.} this myriad of possibilities. The thinness of this common core will become undeniable when we consider all models on the table, but is already very clear when restricting to mainstream candidates. Examples of classes of mainstream particle dark matter candidates, for instance, are weakly interacting massive particles (WIMPs), axions and sterile neutrinos. WIMPs are massive particles that interact via gravity but also via at least one other force (within or outside of the standard model of particle physics) with a strength comparable to or weaker than the weak force. Axions are spin-0 particles---hence, bosons---that solve the strong CP problem in quantum chromodynamics (QCD). Their masses are much smaller than those of WIMPs. Sterile neutrinos interact with gravity but not with any of the fundamental interactions of the standard model (except for experiencing a diluted form of the weak force via mixing with ordinary neutrinos \citep{bertonetait2018}). 

Consider a specific model (including specific values for all parameters) falling into one of these mainstream classes of dark matter candidates, such as the following arbitrary example \citep{cuoco2018}:
\begin{description}
	\item[Example of a (maximally) Thick Concept of Dark Matter:] A single type of supersymmetric WIMP, with zero hypercharge, a mass of 2.8 TeV, an annihilation cross-section into standard model vector bosons of $\langle\sigma v\rangle \sim 10^{-25} cm^3/s$, no (further) self-interactions and a cosmic density corresponding to a contribution of 27\% to the total cosmic mass-energy budget.
\end{description}
It is as clear what it means to be a scientific realist about such a completely specified entity as it is to know what it means to be a scientific realist about, say, the electron in the standard model of particle physics. The problem is that the current empirical evidence strongly underdetermines which, if any, of the vast array of completely specific mainstream candidates is to be paired with $\Lambda$CDM. If those were to differ only in the details, this would not be so bad a problem. However, even in the best case scenario, when we only consider `mainstream' candidates, we would already arrive at something like the following thin concept:
\begin{description}
	\item[The Thin Common Core Concept of Mainstream Dark Matter:] A massive field with a contribution to the total cosmic mass-energy budget of 27\%, thereby being responsible for certain gravity-mediated observables related to structure formation, clusters and galaxies. In case it is a particle, its mass is roughly between $10^{-22} - 10^{13}$ eV.
\end{description} 
Note first that this common core does not require dark matter to be a particle. The current evidence for dark matter consists only of gravity-mediated effects on cosmological and astrophysical observables: galaxy rotation curves, galaxy cluster dynamics, gravitational lensing, and structure formation and the cosmic microwave background. Attempts at direct and indirect detection of the particle nature of dark matter, as well as particle dark matter production at colliders, have (at the time of writing) produced only null results, thereby generating strong upper limits on various hypothetical couplings to standard model particles and of dark matter to itself. Note, thus, that what makes this common core so semantically thin becomes clearest when focusing on all the things that are not included: a unique cosmic density (in case it is a particle); any specific symmetries of the action governing this field, and, relatedly, a list of which non-gravitational interactions---including self-interactions---are experienced by the field, if any(!), as well as corresponding cross-sections.\footnote{The dark matter common core is not only semantically very thin, but also explanatorily and predictively. Part of the explanatory power of dark matter derives from unificatory promises. For instance, supersymmetric WIMP models provide not only a dark matter candidate, but some models also promise to solve the hierarchy problem, the matter-antimatter asymmetry problem, and to unify the various coupling constants of the standard model of particle physics better than the standard model itself does so. Axions solve both the dark matter problem and the strong CP problem. Some sterile neutrinos solve both the dark matter problem and the problem of the massiveness of standard model neutrinos. These specific models kill several birds with one stone, explaining a lot with a little---the common core of DM by itself retains none of this explanatory power. Similarly, it is the specific models that provide predictions and are falsifiable, which is much less so for the mere common core. In a similar vein, the mere common core does not stand a chance at providing a proper explanation (as opposed to post-hoc curve-fitting) of various galactic correlations that modified gravity advocates---the main alternative research programme---emphasise, such as the baryonic Tully-Fisher relation, the mass-discrepancy-acceleration relation, or Renzo's rule. A more detailed analysis and evaluation of the explanatory power of dark matter (and of modified gravity) is left for future work.} 
This common core, thin as it is already, is still a conservative estimate of which dark `matter' candidates should be included as mainstream candidates. Some authors would consider some of the alternative candidates discussed below to also be mainstream candidates; the above is thus a conservative upper bound on the common core of the dark matter concept that is supposed to mesh with the $\Lambda$CDM model, by (partially) filling in the `CDM-slot'.\footnote{It is perhaps not surprising that $\Lambda$CDM exhibits only a semantically thin `CDM-slot'---a ``placeholder'' or ``docking station'' \citep{lehmkuhl2019} that requires filling in---as it is based on Einstein's general relativity (GR) which, according to Lehmkuhl's (\citeyear{lehmkuhl2019}) interpretation of Einstein, is a `hybrid theory'. By this he does \emph{not} mean `hybrid' in the sense of fn.\ref{hybrid}, but rather in the sense of GR being ``fundamental and correct as far as gravity [is] concerned but phenomenological and effective in how it account[s] for matter. As a result, Einstein saw energy-momentum tensors ... in GR as placeholders for a theory of matter not yet delivered'' (p.176).}

The lower bound barely says anything at all:
\begin{description}
	\item[The Thinnest Common Core Concept of Dark `Matter':] Stuff that either contributes 27\% to the total cosmic mass-energy budget or acts as if it does so, thereby being responsible for certain gravity-mediated observables related to structure formation, clusters and galaxies.
\end{description}
`Stuff' is as vague as things can be, but that is exactly because the thinnest common core concept covers almost anything. It may be a type of particle in some regimes, but act instead as phonons, collective excitations of a superfluid Bose-Einstein condensate, in other regimes \citep{berezhiani2015,berezhiani2016}.\footnote{Since the authors conceive of the particle phase as consisting of axionlike particles---which, as bosons, naturally condense below a critical temperature---superfluid dark matter theory is arguably a version of one of the mainstream dark matter candidates.} It need not be a single type of particle, but could be a whole dark QCD-like sector. It need not even consist of (one or several types of) particles in the first place, but could be made up out of primordial black holes. It may be an ``illusion'', created by the gravitational polarization of known matter \citep[p.215]{hajdukovic2011} \citep{blanchet2008}. It could be a modification of gravity \citep{milgrom1983,famaey2012} rather than a new form of massive matter. Critics of modified gravity tend to be more positive about `hybrid theories', which postulate something that acts like matter at cosmological scales and like a modification of gravity at galactic scales.\footnote{\label{hybrid}\citet{berezhiani2015,berezhiani2016,blanchet2008,zhao2008,bruneton2009,li2009,ho2010,ho2011,ho2012,cadoni2018,cadoni2019,scholz2020,skordis2020,martenslehmkuhl1,ferreira2020}.} This may make matters even worse though, as, at best, the `stuff' is both matter and a modifiation of gravity, or, at worst, it becomes unclear whether it is still appropriate to use these categories at all \citep{martenslehmkuhl2}. 
Finally, the answer could be any combination of the above. All that we really know is that luminous matter combined with standard Einsteinian gravity is empirically inadequate. The observed effects are more pronounced than expected from luminous matter plus Einstein's general relativity. We need something more. An extra, mysterious, dark, hidden `entity'---or several such `entities'. Is this something we could be scientific realists about, as of right now?

In all fairness, the alternative solutions in the previous paragraph have been heavily criticised (although it is not at all the case that there has been no criticism of mainstream approaches and $\Lambda$CDM in general\footnote{Some examples are the small-scale challenges \citep{bullock2017,debaerdemaekerboyd2020}, the Bullet Cluster (which, despite usually being put forward as a smoking gun against modified gravity explanations of dark matter data, is potentially also problematic for $\Lambda$CDM) \citep{leekomatsu2010,kraljic2015,asencio2021}, and the tension between different measurements of the  Hubble constant. See also \citep[\S 17.3.2]{kroupa2012}.}). 
The Thinnest Common Core Concept of Dark Matter is really just a loose lower bound. A reasonable appraisal of what is warranted by the available evidence will lie somewhere in between the upper and lower bounds on the semantic common core. But the point stands: the current evidence is all gravity-mediated and severely underdetermines the specific models of dark stuff---the difference not being a matter of mere details, but to the extent that the evidence currently commits us only to a semantically very thin common core notion of dark `matter'.

\todo{
\url{https://inspirehep.net/literature/1691986}
\url{https://arxiv.org/pdf/1701.08720.pdf}
}






\section{Genes, before they were genes} \label{protogenes}

To illustrate issues of scientific realism in the context of such a flimsy and mysterious entity, it will pay off to compare the case at hand to a similar scenario but one with a much longer history. This allows us to peek ahead into what might be an analogous future (\hyperref[discussion]{\S\ref{discussion}}). I believe that doing so should make us humbly realise that the future of dark matter might not be as straightforward and linear as current dark matter realists seem to think is warranted by the evidence, or at the very least that we have some way to go before committing to any substantive realism about dark matter. The scientific concept I have in mind is that of genes, or rather proto-genes, known as Mendelean `factors' or `elements' of inheritance. 

In the 1850s and 60s, the monk Gregor Mendel performed experiments in plant breeding, in order to observe the development of certain traits across generations. He observed a striking pattern of disappearance and reappearance of certain traits, which led him to postulate ``an inherited factor that can be masked and revealed---something passed intact through the apparent loss'' \citep[p.81]{PGS2014}. Eventually this concept developed into what we now know as genes---more on this in \hyperref[discussion]{\S\ref{discussion}}---but even if one is now a realist about genes, one may ask whether realism about factors or proto-genes was (already) justified in the 1860s.

As in the case of dark matter we should start by asking: realism about what exactly? Mendel's factors were as mysterious and flimsy as dark matter is now. At best we had some form of functionalism about proto-genes: ``objects that play a causal role, and [not until] later work [was it] uncovered what actually plays that role'' \citep[p.83]{PGS2014}. But if this role is just a specific cross-generational pattern of traits, this is hardly more informative than Moli\`{e}re's dormitive virtues being responsible for opium making people sleep. It sounds a lot like a mere bookkeeping device for those specific cross-generational patterns of traits, just as the anti-realist logical empiricists defined an electron operationally, partially as a complicated laboratory procedure involving a bubble chamber and observation of a specific pattern of bubbles. It was not at all clear what kind of thing a Mendelean factor was supposed to be: a particle, a cell, a biological substance, a chemical entity or substance, a simple or composite structure, etc? Compare this with the thinnest common core concept of DM leaving open whether it is matter or a modification of the gravitational field. Proto-genes did not (yet) have any ``materiality'' \citep[p.82]{PGS2014} to them: no location, no ontological category, no mechanism. In fact, Mike Buttolph has argued that the late acceptance of Mendel's work is due to it being too simple \citep{challmartens2020}. Even when Johannsen finally coined the term `gene' in 1909, he considered it to be a ``concept ``completely free of any hypothesis'' regarding localization and material constitution'' \citep{SEPgene,johannsen1909}. 
He was a conscious ``agnostic with respect to the material constitution of the genotype and its elements. ....[T]he experimental regime of Mendelian genetics .... did neither require nor allow for any definite supposition about the material structure of the genetic elements'' \citep{SEPgene}.  ``[G]enes were taken as abstract elements of an equally abstract space, whose structure, however, could be explored through the visible and quantifiable outcome of breeding experiments'' \citep{SEPgene}. Even Thomas Morgan, whose group achieved results suggesting that genes lie on chromosomes, noted as late as in his 1933 Nobel Prize lecture that genes could still be fictions, hypothetical units, rather than material things \citep[p.82]{PGS2014} \citep[p.3]{morgan1935} \citep{SEPgene}. This early history is only the beginning of a ``long-standing duality ... between using the word ``gene'' merely to organize talk about observable differences between organisms that show up in breeding experiments in certain ways, and as an attempt to refer to a real hidden object of some kind'' \citep[p.85]{PGS2014}.

It might be interesting to note that some advocates of modified gravity have similarly started referring to Modified Newtonian Dynamics (MOND)---the typical quasi-Newtonian limit that these modifications of gravity aim for---as a mere algorithm that predicts and organises certain galactic observables, rather than as, for instance, a force, or even just a theory \citep{sandersmcgaugh2002}. 

As we will see in \hyperref[discussion]{\S\ref{discussion}}, the subsequent history of discovering what exactly a gene is is rather messy. Regardless, proto-genes and genes in the early days were clearly in the first instance an algorithm for predicting cross-generational patterns of traits, albeit a strikingly effective one indeed. They told us \emph{that} specific observables exhibit a certain pattern, but not \emph{how} or \emph{why}. I contend that it would not have been justified to be a realist about genes \emph{at the time}. Moreover, even if we now had an unambiguous metaphysical picture of what a gene is---a claim that will be critically evaluated in \hyperref[discussion]{\S\ref{discussion}}---this would not retro-actively justify realism about genes back then. 

The common core of the contemporary concept of dark matter is in many ways like this early concept of genes, and for similar reasons of epistemic humility I believe that we cannot (yet) be realists about dark matter. There is of course one glaring disanalogy though: whereas there was a lack of more specific suggestions in those early days for what a gene is, there is currently a whole industry that generates new dark matter models on a daily basis. However, the common core of those models that has been confirmed by the available evidence barely has any substance to it---it is as much of an invisible black box as Mendel's factors.\footnote{Benzer still referred to genes as a black box in the mid-1950s \citep[p.277,284]{holmes2006}.}

\section{The semantic dimension of scientific realism} \label{semrealism}

\begin{figure}
	\centering
	\includegraphics[width=0.85\textwidth]{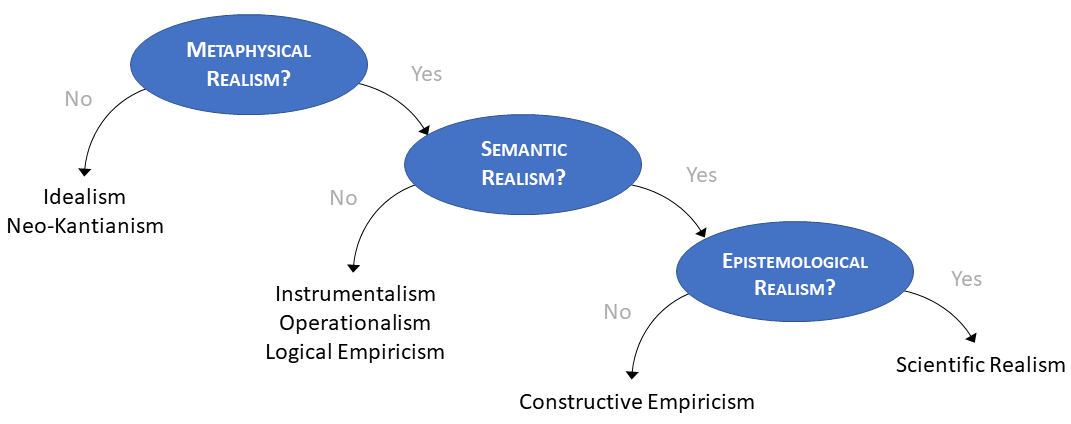}
	\caption{Traditional decision tree for scientific realism vs anti-realisms (or agnosticisms).} \label{fig1}
\end{figure}

This section explains in more detail the type of view---``indefinitely suspended realism'' about dark matter (and early genes)---that I have in mind. There is a variety of forms of scientific anti-realism, or rather degrees of (anti-)realist commitment, of which it is usually assumed that they follow a strict logical order, as per the decision tree in Figure \ref{fig1} \citep{SEPscientificrealism}. 

The first question to ask, that of metaphysical realism, is whether one believes in a mind-independent external world. A negative answer to this question corresponds to the strongest form of anti-realism, metaphysical anti-realism, forms of which are idealism and neo-Kantianism. 

If and only if metaphysical realism is committed to does it make sense to consider the second question. When scientific theories postulate specific theoretical/unobservable terms, say electrons, should we construe such claims as meaningful statements about that mind-independent world (beyond a mere shortcut for specific patterns of observables that stand in some dependence relation---as described by the theory---to those unobservable terms)? Should we take scientific claims about unobservables literally? A negative answer constitutes the second-strongest form of anti-realism, semantic anti-realism, a route famously taken by logical empiricism. 

Supposedly it then only makes sense to ask the final question if one does commit to semantic realism. Roughly speaking, this final question asks whether we should be optimistic or pessimistic about science determining the truth of claims about specific unobservables. More precisely, should we expect science (to be able) to determine whether specific unobservables---literally construed---exist and what their properties are? An affirmative answer corresponds to epistemological realism, which together with the previous two forms of realist commitment constitutes scientific realism (although there is still a choice between being a realist about the entities or only the structure of the theory). Van Fraassen's constructive empiricism famously gives a negative response, epistemological anti-realism, the weakest form of scientific anti-realism \citep{vanfraassen1980}. Science will not give us the truth about unobservables (but that is ok, because that was never the aim of science in the first place according to Van Fraassen).

Where an aspiring dark matter realist thus runs into trouble, given the present-day empirical and theoretical status of dark matter research, is the semantic commitment of scientific realism. The common core concept of dark `matter', especially the lower bound, is so semantically thin, so vacuous, that it barely means anything at all. It is simply not a rich enough concept (yet) for us to be realists about it. If one were to ask a contemporary realist about dark matter what exactly they believe in when they proclaim belief in stuff that acts as if it contributes 27\% to the total cosmic mass-energy budget, and the available evidence does not allow them to give a much more specific answer, then that does not sound like an interesting, substantive species of scientific realism at all. 

What I am thereby \emph{not} saying of course is that we can therefore never have more precise knowledge about this dark stuff, and could therefore never become scientific realists about dark matter. It is crucial to note that the motivation for this selective, as-of-right-now denial of semantic realism, i.e.\ specifically about dark matter, differs completely from what motivated logical empiricists to commit to their specific version of semantic anti-realism. The relevant cornerstone of logical empiricism is the verification principle of meaning. According to this principle, for a scientific statement to be meaningful is for it to be verifiable. Unobservables/ theoretical terms were not considered verifiable, or at least not in a sense that was direct or appropriate enough. Electrons were thus not to be thought of literally, as, say, charged particles. `Electron' is merely a string of symbols, a label that summarises a complicated laboratory procedure resulting in a specific pattern of observables. This principled motivation results in a semantic anti-realism of maximal scope, with respect to types of unobservables and scientific theories as well as with respect to temporal duration. Their semantic anti-realism was thus universal (i.e.~applicable to all unobservables in all theories) and eternal (i.e.~independent of current or future evidence). It is thereby not relevant at all---in contrast to the problems for dark matter realism diagnosed in the current paper---whether theory plus experiment specify exactly what the properties of such theoretical terms would be (e.g.\ in the case of electrons) or fail to do so (e.g.\ in the case of dark matter). Moreover, given such permanent lack of meaning of unobservable terms one can then not even formulate the question of epistemological realism (or perhaps one should say that that question is answered with a trivial no). 

\begin{figure} 
	\centering
	\includegraphics[width=0.85\textwidth]{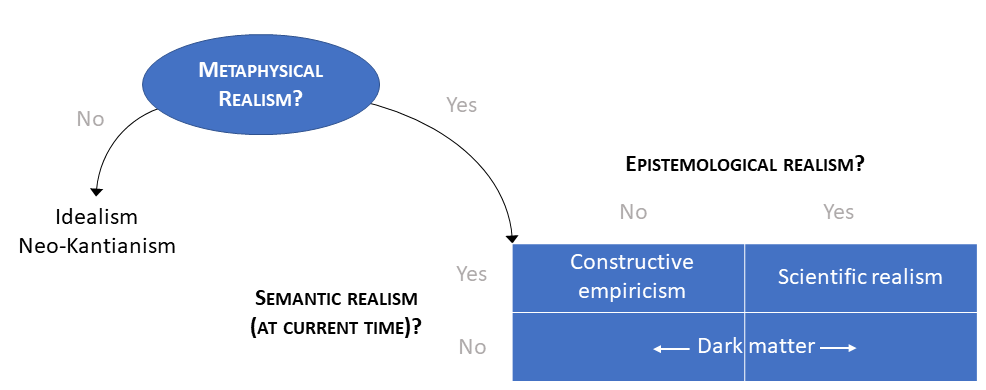}
	\caption{Alternative decision structure for scientific realism vs anti-realisms (or agnosticisms), with the semantic and epistemological components having become independent dimensions of realist commitment. Note that the question of semantic realism is now indexed by the current time.} \label{fig2}
\end{figure}

The indefinite suspension of semantic realism that I advocate in the specific context of dark matter stands in stark contrast to this. There is no principled methodological constraint that underlies the threat to dark matter realism. It merely arises in this specific context due to the strong underdetermination of the common core of the dark `matter' concept by the \emph{currently} available evidence pertaining to this concept. Whereas the universal semantic anti-realism of the logical empiricists was incompatible with epistemological realism---one did not even reach the final question in the decision tree of Figure \ref{fig1}---this maybe-never-but-definitely-not-yet-semantic-realism is logically compatible with both epistemological realism and anti-realism. If future empirical data is capable of pinpointing (almost) exactly what dark `matter' is, that is, if more data could overcome the underdetermination of current models, this would enrich the semantic common core of the dark `matter concept': semantic and epistemological realism would be vindicated simultaneously. The view advocated in this paper is consistent with being an optimist about science being able to narrow down the DM common core, to enlarge its evidence-supported semantic content sufficiently to make it clear what it is that one would be a realist about. Figure \ref{fig2} indicates this alternative logical structure between metaphysical, semantic and epistemological realism.

Note that the view advocated in this paper is thus distinct and independent from Hacking's well-known anti-realism about entities that fall under the purview of astronomy (and presumably also cosmology) \citep{hacking1989},\footnote{For critical responses, see \citep{shapere1993,sandell2010,anderl2016}.}
as I take his view to be a clear instance of epistemological anti-realism even though he states that he is not advocating any particular type of anti-realism. After all, his arguments are based on the lack of the possibility to intervene, the indirectness of the observations, and the problematic use of models. These do not relate to the semantic dimension of realism about the unobservables in question. For instance, with respect to his chosen case study, that of gravitational lenses, the issue is not that it is not clear what type of thing these are, i.e., what the concept of a gravitational lens is. The theory is quite clear on this. The arguments focus on supposed epistemological difficulties related to token gravitational lenses, that is of confirming whether there exists a gravitational lens in a specific region in the universe and what the exact values for this specific object are of the variables that describe the set of possible gravitational lenses. Contrast this with the (partially) astronomical entity that we are after, dark matter: we barely even know what type of thing dark matter is, let alone whether any and, if so, `how many' dark matter `units' there are in a specific region in space. (It should be noted though that if dark matter is a particle, it falls not just under the purview of astronomy and cosmology, but also of particle physics. `Direct detection' or production of these particles on earth would most likely be a way to circumvent Hacking's worries for at least the specific case of dark matter, more amenable to his interventionist criteria for realism.) That the species of anti-realism he advocates is epistemological rather than semantic is confirmed further when he ends his paper by saying that ``[a]strophysics is almost the only human domain where we have profound, intricate knowledge, and in which we can be no more than what [V]an Fraassen calls constructive empiricists'' \citep[p.578]{hacking1989}, i.e.\ semantic realism and epistemological anti-realism. Even more relevant is that he is dismissive of semantic realism/anti-realism debates, considering them to be empty due to their typical lack of attention to the details of science.  Whether that verdict does or does not do justice to logical empiricism, the argument against semantic realism in this paper \emph{is} naturalistic, as it receives its sole motivation from the state of the currently available evidence in cosmology, astrophysics and particle physics. In sum: the indefinite suspension of semantic realism about dark matter in this paper is---although perhaps compatible with it---distinct from Hacking's epistemological anti-realism about all astronomical unobservables.

Up till this point I have been careful to only make a negative claim, i.e.\ that we currently do not have sufficient justification for adopting a specific position, namely semantic realism about dark matter. This seems to leave us with two options for positions that we could \emph{affirm}: ``semantic anti-realism'' or semantic agnosticism about dark matter, i.e.\ the disjunction of semantic realism and ``semantic anti-realism''. If this is indeed the appropriate dilemma (but see below), I sympathise with agnosticism. After all, although there currently isn't sufficient justification for semantic realism and, \emph{a fortiori}, scientific realism about dark matter, we do not have sufficient justification for its negation either---after all, we have good justification for believing that postulating only luminous matter (plus the assumption of Einstein's theory of general relativity) will not suffice to explain or even just account for the gravity-mediated cosmological observables mentioned in Section \ref{DM}.

Agnosticism is `right in between' realism and anti-realism. Shapere seems to disagree, considering agnosticism to be potentially much closer to realism than to anti-realism, when he proposes what seems like a mixture of agnosticism and realism, in the specific context of gravitational lenses \citep{shapere1993}. However, the agnostic aspect of his proposal is in fact much less radical than agnosticism about realism vs anti-realism, since he means that we can (and should) be realists about the existence of gravitational lenses even though, for various reasons, we have to remain agnostic about (the exact value of) a few of their properties. In fact, it is not radical at all, but reflects nothing more than the norm, made inevitable by the limited accuracy of any experiment: of no entities that we are realists about do we know all their properties infinitely precisely. 

In the spirit of Shapere, one may even try to argue against the main thesis of this paper as follows. Although the semantic content of the common core concept of dark `matter' is indeed very small, that is still more than nothing. Should we then not conclude a very meagre form of (semantic) realism, rather than anti-realism or agnosticism? I disagree. Such a supposed form of minimal realism is much closer to anti-realism or agnosticism than it is to being a (semantic) realist about, say, electrons. If one does not know whether one is believing in either, say, primordial black holes or an ultralight boson with a De Broglie wavelength of the order of 1 kpc\footnote{Fuzzy dark matter \citep{hui2017}.} (or one of several other distinct options), then asserting the belief in the disjunction of these radically distinct options and considering that to be a form of realism is much worse than, say, claiming oneself to be a realist about electrons in virtue of believing in muons.

To sum up: realism with a pinch of agnosticism/uncertainty about the exact properties of the entity in question (for instance about gravitational lenses) is more than fine, but in the case of dark matter the uncertainty dominates to the extent that it becomes unrecognisable as honest realism. This suggests instead the plausibility of pure overall agnosticism about semantic realism vs ``semantic anti-realism'' about dark matter. There may well be other case studies that fall in a grey area, i.e.\ the thickness of the semantic common core of the concept in question is right in between that of dark matter and of electrons, such that it may be tricky to decide whether that common core is semantically thick enough to satisfy the semantic component of realism. Although I wouldn't necessarily have anything in general to say about such grey cases,\footnote{One option could be to default to agnosticism, another option could be to bring in degrees of belief in realism.} their existence would not imply that there are no fairly black or white cases, which I indeed take dark matter to be an instance of.

All that being said, it's important to note the following subtlety. Although realism, agnosticism and anti-realism may well be the appropriate and complete list of options as far as the metaphysical and epistemological components of realism as well as scientific realism \emph{tout court} are concerned, the semantic component is special. What is at stake is, among other things, exactly whether the unobservable entity in question is conceptually substantive/well-defined/meaningful enough, and 	\emph{thereby also} whether claims \emph{both} about the existence of this entity (i.e.\ realism) as well as about its non-existence (i.e.\ anti-realism) are meaningful. If the term `dark matter'---in the sense of one of the semantic common cores in Section \ref{DM}---is not very meaningful (yet), claiming that dark matter exists or claiming that dark matter does not exist (or claiming the disjunction of these two claims) is equally vacuous. Satisfying the semantic component---or at least that aspect of the semantic component that concerns the concept under consideration being well-defined enough---thus seems to be a \emph{precondition} for being able to formulate the scientific realism vs.\ anti-realism disagreement. 

A potential response is as follows. Although it was pointed out above that it would not help the semantic realist to replace the common core by a disjunction of many (known)\footnote{Another worry here is that there may be unconceived alternatives; there is no guarantee that, even if we end up finding dark matter, it will be one of the options we have already thought of.} dark matter models, things look better if one were to cash out anti-realism as the belief that it is not the case that that disjunction exists, which would be equivalent to believing in the conjunction `black hole dark matter does not exist AND fuzzy dark matter does not exist AND ... etc.'; this would be a coherent position. However coherent, there is no indication that this anti-realist position about dark matter is currently justified. If one were to insist on affirming a specific position regarding dark matter realism---but recall that the upshot of this paper is primarily the negative claim that it is too early to be a (semantic) realist about dark matter---we're thus back at (semantic) agnosticism.

\section{Genes} \label{discussion}

The main point of this paper has now been made---the current evidence does not (yet) justify selective semantic realism about dark `matter', and \emph{a fortiori} no full-blown scientific realism about dark matter either. One may wish to retort however that we know for certain that luminous matter plus Einstein's laws of gravity cannot be the full or correct story. We know that something is missing. Do we not have good reason to believe that physics will narrow down the common core concept of dark `matter' in the foreseeable future? And if so, does this guarantee not allow us to already be scientific realists, including semantic realists, about dark matter right now? Before responding to these questions, let us continue the history of genes up until current times \citep{PGS2014,SEPgene,SEPgenetics}. Be prepared for a messy story---too messy to vindicate contemporary realism about genes and arguably even messy enough to justify contemporary anti-realism about genes.





Genetics has advanced immensely since Mendel, initially corroborating the simplicity of Mendel's factors but eventually revealing a concept that is much more complex than expected. After Mendel, genes could be accessed more directly, not merely through breeding experiments. Classical genetics focused on the chemical nature of the gene, conceiving of each gene as being responsible for the production of a single enzyme---a protein that is a biological catalyst for chemical reactions in the cell. Classical genetics transitioned to molecular genetics in the 1950s, with as highlight the 1953 DNA model of Watson and Crick and the subsequent uncovering, in the 60s, of the genetic code. This is a mapping from stretches of DNA to protein structure. The concept of a ``gene'' was eventually identified with Benzer's concept ``cistron'' \citep{benzer1957}, the DNA stretch that codes for the structure of a single protein molecule. At this point it may seem that we have a vindication of Mendel's factors of inheritance, a tangible and semantically thick concept of a gene that is a precisification, a filling in, of the original idea. 

However, the cistron is merely a `unit of function' of DNA, a `unit' that is, in various senses, neither neatly localised, nor necessarily temporally stable, nor defined uniquely or without vagueness, nor a complete, context-independent explanation of observed breeding patterns. It is not clear that we should talk about a unified concept, a unit, at all. Consider first locality and stability. A single cistron need not be contiguous---it can be interspersed with non-coding stretches of DNA called introns \citep{berget1977,chow1977,SEPgenetics}, 
or it can even correspond to stretches of code from two discrete organisational collections of DNA called chromosomes (via a process referred to as RNA trans-splicing).
Conversely, multiple cistrons need not be (fully) separated, as their associated DNA stretches can partially overlap \citep{barrell1976,SEPgenetics}
or one code can even be embedded in the other. 
More radical even than RNA trans-splicing---where the stretches of code from two chromosomes are used but the chromosomes are left intact---is recombination of genes. During meiosis chromosomes can cross over; they break and exchange genetic material in a way that does not respect cistron boundaries. The unit of recombination that is ``\emph{passed on intact} is not the same as the [functional] \emph{unit that makes a protein}'' \citep[p.84]{PGS2014}! This lead Benzer to conclude by the mid 1950s that the word `gene' had become a ``dirty word'' \citep[p.285]{holmes2006} \citep{SEPgene}. For similar reasons it is also not clear that it is the correct evolutionary unit that is selected for, with the cross-generational stability that Mendel expected. Although evolution is often even defined in terms of gene frequency, with the gene being passed on as a discrete unit, this is arguably an idealisation \citep[\S 6.3]{PGS2014}. A gene is arguably both too large---cistrons are broken when gene recombination occurs via sex or meiosis---and too small, as larger stretches of genetic code (of various sizes!) are usually selected for, i.e.\ genes tend to be evolutionarily relevant in the context of other genes rather than by themselves. There is no unique length of a DNA stretch that is passed on intact across generations, contra the spirit of Mendel's original ideas.


Consider now the ambiguities in the definition of the concept of a gene, as well as its merely partial, context-dependent role in accounting for observables. The complexity of modern genetics has lead some authors to simultaneously recognize multiple gene concepts.
Moss (\citeyear{moss2002}) modestly starts off with two concepts: gene-P (an instrumentalist definition---anything in the genome that relates to predicting the observable phenotype, without requiring a one-to-one relation between genes and traits---somewhat resembling early genes) and gene-D (a specific molecular sequence). Griffiths and Stotz (\citeyear{griffithsstotz2013}) employ three or four concepts.\footnote{For other examples of multiple gene concepts, see \citep{foxkeller2002,baetu2011}.} Before that, Sterelny and Griffiths (\citeyear[p.133]{sterelnygriffiths1999}) even went as far as claiming that the concept of `gene' is context dependent, it being used as a ``floating label'' for any bit of DNA that is of interest. 

Moreover, the gene-D is by itself indeterminate with respect to phenotype; its interaction with a plethora of other developmental resources is what determines the observable traits \citep{SEPgene,SEPgenetics}. For instance, molecules binding to `promoters' located near the cistron can determine whether that stretch of code is or is not transcribed into RNA, and raw RNA transcripts can then be alternatively spliced by trans-acting repressors and activators into various finished RNA transcripts for protein synthesis. 
These two and a myriad of other factors determine whether and how the gene contributes to the synthesis of proteins and is eventually responsible for traits. Interestingly, the white-eye `gene' in fruit flies that made Morgan's group (see above) famous turned out to be a mutation in a promoter instead \citep{PGS2014}. Explanation of cross-generational patterns of a trait has become context-dependent, a holistic story, rather than being reducible to a single gene, an ontologically privileged ``atom of inheritance'' \citep[p.85]{PGS2014}. 

Add to this multiplicity of concepts of genes and the holistic, complex nature of modern genetics that there does not seem to be a unique genetic unit of evolution, and we realise that `genes' are, at the very least, ``more indefinite and blurry entities than had been supposed'' \citep[p.97]{PGS2014}. More strongly put, ``[g]enomes are more organzied objects, and their partition into genes more artificial, than the classic models suppose'' \citep[p.99]{PGS2014}---a deliberate simplification. ``There is a fact of the matter about the structure of DNA, but there is no single fact about the matter what the gene is''.\footnote{Richard Burian (\citeyear[p.37]{burian1985}), as quoted by Fox Keller (\citeyear[p.66]{foxkeller2002}).} This Mendel-inspired view of a hereditary atom, despite being the driver of success in the early 20th century, might now have become ``a hindrance to our [further] understanding'', ``a concept past its time''\footnote{William Gelbart (\citeyear[p.660]{gelbart1998}), as quoted by Fox Keller (\citeyear[p.67-8]{foxkeller2002}).} that is ``no longer useful''\footnote{Peter Portin (\citeyear[p.208]{portin1993}), as quoted by Fox Keller (\citeyear[p.67]{foxkeller2002}).}.

I am tempted to conclude that we should, at the current time, be anti-realists about genes. Even if one does not agree with this, it is fair to say that the current understanding of the concept of gene has not (just) filled in Mendel's vague concept, but at least significantly augmented it and to an extent replaced it \citep{PGS2014,hull1974}. We may wonder whether we would talk about genes at all, if not for the route via Mendel and Morgan \citep{PGS2014}. Would Mendel recognise the current concept as being in the spirit of what he had in mind, or consider it an unconceived alternative? Even if so, if one were a realist about Mendel's factors in the early days, one would not have known that anything like this modern notion of `gene' is what one was believing in. The Mendelian promise of a unified concept has not born out; `gene' is still, or perhaps again, a semantically thin concept, at best.

Let us return to the questions at the start of this section. Several lessons can be learnt from the history of genes, all revolving around the cautionary theme that we should not count our chickens before they hatch. We may be in situations where we are certain that the ontology of our current theory or theories is not the whole story, or at least not the correct story, but without sufficient knowledge of what is missing we cannot (yet) be realists about the missing entity. To be a realist about the unknown is to not be a realist. This is most obvious in cases where our reasons for believing that the current story is incomplete are purely theoretical. That the standard model of particle physics has what many consider to be internal explanatory gaps, such as a lack of explanation of there being three generations of quark pairs and lepton pairs, does not by itself give any knowledge of that what is supposedly missing. Knowing that quantum theory and general relativity do not mesh well together is not sufficient to know what is missing or what needs to be changed. The cases of dark matter and genes are slightly different, slightly better, in that we have some empirical patterns to go on: cross-generational traits in breeding experiments, galaxy rotation curves, etc. However, we have seen that these initial hints from nature might falsely suggest a simple solution, a simple concept, realism about which would seem to solve the problem. Semantic thinness should not be mistaken for definite simplicity. 

There is no guarantee that we will be able to make the semantic common core concept of dark `matter' more precise, more definite, less blurry, less vague. Even if we do manage to obtain more knowledge, that knowledge may reveal a concept that is more complex, less material, and more context-dependent than many of our current models suggest. Similarly to the case for genes, the locality of dark matter may not be as traditionally expected---the location of fuzzy dark matter \citep{hui2017}, with its De Broglie wavelength of the order of 1kpc, is highly indefinite; Verlinde's entropic gravity designed to mimic dark matter effects is non-local due to its holographic nature \citep{verlinde2017}. A cistron is in the first instance defined functionally, rather than materially, as an entity in a well-defined, contiguous region of space/DNA. The MOND-formalism is sometimes also viewed, in the first instance, as a functional role, an algorithm, rather than a modification of gravity or a dark matter particle. 

Similarly to the case for genes, we may require multiple types of dark matter. After all, the luminous matter sector contains a whole zoo of types of particles, so why would the energetically dominant dark matter sector of the universe not consist of multiple components? 

Similarly to the case for genes, advocates of dark `matter' indeed being matter rather than a modification of gravity tend to expect that the galactic correlations that MOND is known to be able to explain as well as several `small-scale challenges' \citep{bullock2017,debaerdemaekerboyd2020} are in fact due to a complex, messy interaction of dark matter and luminous/baryonic matter. Context matters. 

Similarly to the case for genes, hybrid theories (fn.\ref{hybrid}) suggest, at best, that there are entities that are both matter and spacetime or that are not (conceptually) stable over time (but switch, say, from being an aspect of gravity/spacetime when in galaxies to being dark matter when outside of galaxies) \citep{martenslehmkuhl1}. At worst, they may indicate a blurring of these traditional concepts, or even their inapplicability and need for replacement \citep{martenslehmkuhl2}. It is not guaranteed that once we find dark `matter' it will be recognisable as such by, say, Zwicky, who first considered dark matter in the 1930s, in the context of galaxy clusters, or by the observers of galaxy rotation curves in the 1970s.

All these similarities consider merely conceived alternatives to mainstream dark matter candidates. Unconceived alternatives might make matters worse, just as the intricacies of the current concept of a gene which were very much unconceived of by Mendel. 
  
Finally, even if the future of dark matter research vindicates a simple dark matter model that is close to the concept as originally envisaged, this still does not imply that we are currently  justified in being (semantic) realists about dark matter \emph{already}.

\section{Conclusion} \label{conclusion}

According to the standard model of cosmology, $\Lambda$CDM, the mass-energy budget of the current stage of the universe is not dominated by the luminous matter that we are familiar with, but instead by dark matter (and dark energy). However tempting it may be to adopt scientific realism about dark matter, I advocate caution. Empirical data barely constrain the dark `matter' properties---it is in fact not even certain that it is a form of matter. In light of this thin semantic content \emph{as of right now}, I have argued against (already) adopting semantic realism about dark matter, drawing upon the lessons learnt from the analogous history of the concept of a gene. This denial of the semantic component of dark matter realism being satisfied is thus different from Hacking's anti-realism about all astronomical entities, which is of an epistemological nature. It also differs from the semantic anti-realism of logical empiricism, in that it is naturalistic, in such a way that it does not automatically apply to all other unobservables (or even just to all other astronomical unobservables, as with Hacking's anti-realism) and that it might only be of a temporary nature. We may one day narrow down what dark 'matter' is, but for now we remain in the dark. 

\section*{Acknowledgments}

I would like to acknowledge support from the DFG (Deutsche Forschungsgemeinschaft) Research Unit ``The Epistemology of the Large Hadron Collider'' (grant FOR 2063). Within this research unit I am particularly indebted to the other members of the `LHC, Dark Matter \& Modified Gravity' project team---Miguel \'{A}ngel Carretero Sahuquillo, Michael Kr\"{a}mer, Dennis Lehmkuhl and Erhard Scholz---for invaluable and extensive discussions. I would furthermore like to thank Josh Eisenthal, Jamee Elder, Alison Feder, Sjang ten Hagen, Martin King, Dennis Lehmkuhl, Tushar Menon, Noah Stemeroff and the Lichtenberg Group for History and Philosophy of Physics (University of Bonn) for invaluable discussions and comments on earlier drafts.



\bibliography{dmrealism}

\begin{thebibliography}{65}
\providecommand{\natexlab}[1]{#1}
\providecommand{\url}[1]{\texttt{#1}}
\expandafter\ifx\csname urlstyle\endcsname\relax
  \providecommand{\doi}[1]{doi: #1}\else
  \providecommand{\doi}{doi: \begingroup \urlstyle{rm}\Url}\fi

\bibitem[Anderl(2016)]{anderl2016}
S.~Anderl.
\newblock Astronomy and astrophysics.
\newblock In P.~Humphreys, editor, \emph{The Oxford Handbook of Philosophy of
  Science}. 2016.

\bibitem[Asencio et~al.(2021)Asencio, Banik, and Kroupa]{asencio2021}
E.~Asencio, I.~Banik, and P.~Kroupa.
\newblock {A massive blow for $\Lambda$CDM -- the high redshift, mass, and
  collision velocity of the interacting galaxy cluster El Gordo contradicts
  concordance cosmology}.
\newblock \emph{Monthly Notices of the Royal Astronomical Society},
  500\penalty0 (4):\penalty0 5249--5267, 11 2021.
\newblock ISSN 0035-8711.
\newblock \doi{10.1093/mnras/staa3441}.
\newblock URL \url{https://doi.org/10.1093/mnras/staa3441}.

\bibitem[Baetu(2011)]{baetu2011}
T.~M. Baetu.
\newblock A defense of syntax-based gene concepts in postgenomics: Genes as
  modular subroutines in the master genomic program.
\newblock \emph{Philosophy of Science}, 78\penalty0 (5):\penalty0 712--723,
  2011.

\bibitem[{Barrell} et~al.(1976){Barrell}, {Air}, and {Hutchinson
  III}]{barrell1976}
B.~{Barrell}, G.~M. {Air}, and C.~A. {Hutchinson III}.
\newblock Overlapping genes in bacteriophage {$\Phi$X174}.
\newblock \emph{Nature}, 264\penalty0 (5581):\penalty0 34--41, 1976.
\newblock \doi{10.1038/264034a0}.

\bibitem[Benzer(1957)]{benzer1957}
S.~Benzer.
\newblock The elementary units of heredity.
\newblock In \emph{A Symposium on the Chemical Basis of Heredity}. Baltimore:
  Johns Hopkins University Press, 1957.

\bibitem[Berezhiani and Khoury(2015)]{berezhiani2016}
L.~Berezhiani and J.~Khoury.
\newblock Theory of dark matter superfluidity.
\newblock \emph{Phys. Rev. D}, 92:\penalty0 103510, Nov 2015.
\newblock \doi{10.1103/PhysRevD.92.103510}.
\newblock URL \url{https://link.aps.org/doi/10.1103/PhysRevD.92.103510}.

\bibitem[Berezhiani and Khoury(2016)]{berezhiani2015}
L.~Berezhiani and J.~Khoury.
\newblock Dark matter superfluidity and galactic dynamics.
\newblock \emph{Physics Letters B}, 753:\penalty0 639--643, 2016.

\bibitem[{Berget} et~al.(1977){Berget}, Moore, and {Sharp}]{berget1977}
S.~M. {Berget}, C.~Moore, and P.~A. {Sharp}.
\newblock Spliced segments at the 5' terminus of adenovirus 2 late mrna.
\newblock \emph{Proceedings of the National Academy of Sciences}, 74\penalty0
  (8):\penalty0 3171--3175, 1977.
\newblock \doi{10.1073/pnas.74.8.3171}.

\bibitem[Bertone and Tait(2018)]{bertonetait2018}
G.~Bertone and T.~M. Tait.
\newblock A new era in the search for dark matter.
\newblock \emph{Nature}, 562\penalty0 (7725):\penalty0 51--56, 2018.

\bibitem[Blanchet and Le~Tiec(2008)]{blanchet2008}
L.~Blanchet and A.~Le~Tiec.
\newblock Model of dark matter and dark energy based on gravitational
  polarization.
\newblock \emph{Phys. Rev. D}, 78:\penalty0 024031, 7 2008.
\newblock \doi{10.1103/PhysRevD.78.024031}.
\newblock URL \url{https://link.aps.org/doi/10.1103/PhysRevD.78.024031}.

\bibitem[Bruneton et~al.(2009)Bruneton, Liberati, Sindoni, and
  Famaey]{bruneton2009}
J.-P. Bruneton, S.~Liberati, L.~Sindoni, and B.~Famaey.
\newblock Reconciling {MOND} and dark matter?
\newblock \emph{Journal of Cosmology and Astroparticle Physics}, 2009\penalty0
  (03):\penalty0 021--021, mar 2009.
\newblock \doi{10.1088/1475-7516/2009/03/021}.
\newblock URL \url{https://doi.org/10.1088%2F1475-7516%2F2009%2F03%2F021}.

\bibitem[Bullock and Boylan-Kolchin(2017)]{bullock2017}
J.~S. Bullock and M.~Boylan-Kolchin.
\newblock Small-scale challenges to the {$\Lambda$CDM} paradigm.
\newblock \emph{Annual Review of Astronomy and Astrophysics}, 55\penalty0
  (1):\penalty0 343--387, 2017.
\newblock \doi{10.1146/annurev-astro-091916-055313}.
\newblock URL \url{https://doi.org/10.1146/annurev-astro-091916-055313}.

\bibitem[{Burian}(1985)]{burian1985}
R.~{Burian}.
\newblock \emph{On conceptual change in biology: the case of the gene}, pages
  21--42.
\newblock Cambridge: MIT Press, 1985.

\bibitem[Cadoni and Tuveri(2019)]{cadoni2019}
M.~Cadoni and M.~Tuveri.
\newblock Galactic dynamics and long-range quantum gravity.
\newblock \emph{Phys. Rev. D}, 100:\penalty0 024029, Jul 2019.
\newblock \doi{10.1103/PhysRevD.100.024029}.
\newblock URL \url{https://link.aps.org/doi/10.1103/PhysRevD.100.024029}.

\bibitem[Cadoni et~al.(2018)Cadoni, Casadio, Giusti, M\"{u}ck, and
  Tuveri]{cadoni2018}
M.~Cadoni, R.~Casadio, A.~Giusti, W.~M\"{u}ck, and M.~Tuveri.
\newblock Effective fluid description of the dark universe.
\newblock \emph{Physics Letters B}, 776:\penalty0 242--248, 2018.
\newblock ISSN 0370-2693.
\newblock \doi{https://doi.org/10.1016/j.physletb.2017.11.058}.
\newblock URL
  \url{http://www.sciencedirect.com/science/article/pii/S0370269317309504}.

\bibitem[Chakravartty(2017)]{SEPscientificrealism}
A.~Chakravartty.
\newblock {Scientific Realism}.
\newblock In E.~N. Zalta, editor, \emph{The {Stanford} Encyclopedia of
  Philosophy}. Metaphysics Research Lab, Stanford University, summer 2017
  edition, 2017.

\bibitem[{Chall} and {Martens}(2020)]{challmartens2020}
C.~{Chall} and N.~C. {Martens}.
\newblock Simplicity in the sciences and humanities: Report on the {Bonn}
  ``simplicities and complexities'' conference.
\newblock \emph{Journal for General Philosophy of Science}, 51:\penalty0
  491--497, 2020.
\newblock \doi{10.1007/s10838-020-09499-2}.

\bibitem[{Chow} et~al.(1977){Chow}, {Gelinas}, {Broker}, and
  {Roberts}]{chow1977}
L.~T. {Chow}, R.~E. {Gelinas}, T.~R. {Broker}, and R.~J. {Roberts}.
\newblock An amazing sequence arrangement at the 5' ends of adenovirus 2
  messenger rna.
\newblock \emph{Cell}, 12\penalty0 (1):\penalty0 1--8, 1977.
\newblock \doi{10.1016/0092-8674(77)90180-5}.

\bibitem[Cuoco et~al.(2018)Cuoco, Heisig, Korsmeier, and Kr\"{a}mer]{cuoco2018}
A.~Cuoco, J.~Heisig, M.~Korsmeier, and M.~Kr\"{a}mer.
\newblock Constraining heavy dark matter with cosmic-ray antiprotons.
\newblock \emph{Journal of Cosmology and Astroparticle Physics}, 2018\penalty0
  (04):\penalty0 004--004, apr 2018.
\newblock \doi{10.1088/1475-7516/2018/04/004}.
\newblock URL \url{https://doi.org/10.1088%2F1475-7516%2F2018%2F04%2F004}.

\bibitem[{De Baerdemaeker} and Boyd(2020)]{debaerdemaekerboyd2020}
S.~{De Baerdemaeker} and N.~M. Boyd.
\newblock Jump ship, shift gears, or just keep on chugging: Assessing the
  responses to tensions between theory and evidence in contemporary cosmology.
\newblock \emph{Studies in History and Philosophy of Science Part B: Studies in
  History and Philosophy of Modern Physics}, 72:\penalty0 205--216, 2020.
\newblock ISSN 1355-2198.
\newblock \doi{https://doi.org/10.1016/j.shpsb.2020.08.002}.
\newblock URL
  \url{https://www.sciencedirect.com/science/article/pii/S1355219820301088}.

\bibitem[{Di Luzio} et~al.(2020){Di Luzio}, Giannotti, Nardi, and
  Visinelli]{diluzio2020}
L.~{Di Luzio}, M.~Giannotti, E.~Nardi, and L.~Visinelli.
\newblock The landscape of {QCD} axion models.
\newblock \emph{Physics Reports}, 870:\penalty0 1--117, 2020.
\newblock ISSN 0370-1573.
\newblock \doi{https://doi.org/10.1016/j.physrep.2020.06.002}.
\newblock URL
  \url{https://www.sciencedirect.com/science/article/pii/S0370157320302477}.

\bibitem[Famaey and McGaugh(2012)]{famaey2012}
B.~Famaey and S.~S. McGaugh.
\newblock Modified newtonian dynamics {(MOND):} observational phenomenology and
  relativistic extensions.
\newblock \emph{Living Reviews in Relativity}, 15\penalty0 (1):\penalty0 10,
  Sep 2012.
\newblock ISSN 1433-8351.
\newblock \doi{10.12942/lrr-2012-10}.
\newblock URL \url{https://doi.org/10.12942/lrr-2012-10}.

\bibitem[{Ferreira}(2021)]{ferreira2020}
E.~G. {Ferreira}.
\newblock Ultra-light dark matter.
\newblock \emph{Astron Astrophys Rev}, 29\penalty0 (7), 2021.
\newblock \doi{10.1007/s00159-021-00135-6}.
\newblock URL \url{https://arxiv.org/abs/2005.03254}.

\bibitem[{Fox Keller}(2002)]{foxkeller2002}
E.~{Fox Keller}.
\newblock \emph{The Century of the Gene}.
\newblock Harvard University Press, 2002.

\bibitem[Gelbart(1998)]{gelbart1998}
W.~Gelbart.
\newblock Data bases in genomic research.
\newblock \emph{Science}, 282:\penalty0 659--661, 1998.
\newblock \doi{10.1126/science.282.5389.659}.

\bibitem[Godfrey-Smith(2014)]{PGS2014}
P.~Godfrey-Smith.
\newblock \emph{Philosophy of Biology}.
\newblock Princeton University Press, 2014.

\bibitem[{Griffiths} and {Stotz}(2013)]{griffithsstotz2013}
P.~{Griffiths} and K.~{Stotz}.
\newblock \emph{Genetics and Philosophy: An Introduction}.
\newblock Cambridge, MA: Harvard University Press, 2013.

\bibitem[Hacking(1989)]{hacking1989}
I.~Hacking.
\newblock Extragalactic reality: The case of gravitational lensing.
\newblock \emph{Philosophy of Science}, 56\penalty0 (4):\penalty0 555--581,
  1989.

\bibitem[{Hajdukovic}(2011)]{hajdukovic2011}
D.~S. {Hajdukovic}.
\newblock Is dark matter an illusion created by the gravitational polarization
  of the quantum vacuum?
\newblock \emph{Astrophysics and Space Science}, 334:\penalty0 215--218, 2011.

\bibitem[Ho et~al.(2011)Ho, Minic, and Ng]{ho2011}
C.~Ho, D.~Minic, and Y.~Ng.
\newblock Quantum gravity and dark matter.
\newblock \emph{Gen Relativ Gravit}, pages 2567--2573, 2011.
\newblock \doi{10.1007/s10714-011-1200-z}.

\bibitem[Ho et~al.(2010)Ho, Minic, and Ng]{ho2010}
C.~M. Ho, D.~Minic, and Y.~J. Ng.
\newblock Cold dark matter with {MOND} scaling.
\newblock \emph{Physics Letters B}, 693\penalty0 (5):\penalty0 567 -- 570,
  2010.
\newblock ISSN 0370-2693.
\newblock \doi{https://doi.org/10.1016/j.physletb.2010.09.008}.
\newblock URL
  \url{http://www.sciencedirect.com/science/article/pii/S0370269310010592}.

\bibitem[Ho et~al.(2012)Ho, Minic, and Ng]{ho2012}
C.~M. Ho, D.~Minic, and Y.~J. Ng.
\newblock Dark matter, infinite statistics, and quantum gravity.
\newblock \emph{Phys. Rev. D}, 85:\penalty0 104033, May 2012.
\newblock \doi{10.1103/PhysRevD.85.104033}.
\newblock URL \url{https://link.aps.org/doi/10.1103/PhysRevD.85.104033}.

\bibitem[{Holmes}(2006)]{holmes2006}
F.~L. {Holmes}.
\newblock \emph{Reconceiving the Gene: Seymour Benzer's Adventures in Phage
  Genetics}.
\newblock New Haven: Yale University Press, 2006.

\bibitem[Hui et~al.(2017)Hui, Ostriker, Tremaine, and Witten]{hui2017}
L.~Hui, J.~P. Ostriker, S.~Tremaine, and E.~Witten.
\newblock Ultralight scalars as cosmological dark matter.
\newblock \emph{Phys. Rev. D}, 95:\penalty0 043541, Feb 2017.
\newblock \doi{10.1103/PhysRevD.95.043541}.
\newblock URL \url{https://link.aps.org/doi/10.1103/PhysRevD.95.043541}.

\bibitem[Hull(1974)]{hull1974}
D.~Hull.
\newblock \emph{Philosophy of Biological Science}.
\newblock Englewood Cliffs, NJ: Prentice-Hall, 1974.

\bibitem[Jacquart(2021)]{jacquartcompanion}
M.~Jacquart.
\newblock Dark matter and dark energy.
\newblock In \emph{The Routledge Companion to Philosophy of Physics}. 2021.

\bibitem[Johannsen(1909)]{johannsen1909}
W.~Johannsen.
\newblock \emph{Elemente der exakten erblichkeitslehre: Deutsche wesentlich
  erweiterte ausgabe in f\"{u}nfundzwanzig vorlesungen}.
\newblock 1909.

\bibitem[King(1977)]{king1977}
I.~King.
\newblock Galaxies and their populations---the view on a cloudy day.
\newblock In B.~{Tinsley} and R.~{Larson}, editors, \emph{Proceedings of a
  Conference on The Evolution of Galaxies and Stellar Populations, May 1997},
  pages 1--17 (239,278). New Haven: Yale University Observatory, 1977.

\bibitem[Kraljic and Sarkar(2015)]{kraljic2015}
D.~Kraljic and S.~Sarkar.
\newblock How rare is the bullet cluster (in a {$\Lambda$CDM} universe)?
\newblock \emph{Journal of Cosmology and Astroparticle Physics}, 2015.
\newblock \doi{10.1088/1475-7516/2015/04/050}.

\bibitem[Kroupa(2012)]{kroupa2012}
P.~Kroupa.
\newblock The dark matter crisis: Falsification of the current standard model
  of cosmology.
\newblock \emph{Publications of the Astronomical Society of Australia},
  29\penalty0 (4):\penalty0 395--433, 2012.
\newblock \doi{10.1071/AS12005}.

\bibitem[Lee and Komatsu(2010)]{leekomatsu2010}
J.~Lee and E.~Komatsu.
\newblock Bullet cluster: a challenge to {$\Lambda$CDM} cosmology.
\newblock \emph{The Astrophysical Journal}, 718\penalty0 (1), 2010.

\bibitem[Lehmkuhl(2019)]{lehmkuhl2019}
D.~Lehmkuhl.
\newblock General relativity as a hybrid theory: The genesis of {Einstein's}
  work on the problem of motion.
\newblock \emph{Studies in History and Philosophy of Science Part B: Studies in
  History and Philosophy of Modern Physics}, 67:\penalty0 176--190, 2019.
\newblock ISSN 1355-2198.
\newblock \doi{https://doi.org/10.1016/j.shpsb.2017.09.006}.
\newblock URL
  \url{https://www.sciencedirect.com/science/article/pii/S1355219817301314}.

\bibitem[Li and Zhao(2009)]{li2009}
B.~Li and H.~Zhao.
\newblock Environment-dependent dark sector.
\newblock \emph{Phys. Rev. D}, 80:\penalty0 064007, Sep 2009.
\newblock \doi{10.1103/PhysRevD.80.064007}.
\newblock URL \url{https://link.aps.org/doi/10.1103/PhysRevD.80.064007}.

\bibitem[{Martens} and {Lehmkuhl}(2020{\natexlab{a}})]{martenslehmkuhl1}
N.~C.~M. {Martens} and D.~{Lehmkuhl}.
\newblock Dark matter = modified gravity? {Scrutinising} the spacetime--matter
  distinction through the modified gravity/ dark matter lens.
\newblock \emph{Studies in History and Philosophy of Modern Physics},
  72:\penalty0 237--250, 2020{\natexlab{a}}.

\bibitem[{Martens} and {Lehmkuhl}(2020{\natexlab{b}})]{martenslehmkuhl2}
N.~C.~M. {Martens} and D.~{Lehmkuhl}.
\newblock Cartography of the space of theories: an interpretational chart for
  fields that are both (dark) matter and spacetime.
\newblock \emph{Studies in History and Philosophy of Modern Physics},
  72:\penalty0 217--236, 2020{\natexlab{b}}.

\bibitem[{Martens} et~al.(forthcoming){Martens}, {{Carretero Sahuquillo}},
  Scholz, Lehmkuhl, and Kr\"{a}mer]{dmeditorialb}
N.~C.~M. {Martens}, M.~A. {{Carretero Sahuquillo}}, E.~Scholz, D.~Lehmkuhl, and
  M.~Kr\"{a}mer.
\newblock Integrating dark matter, modified gravity, and the humanities.
\newblock \emph{Studies in History and Philosophy of Science}, forthcoming.

\bibitem[{Milgrom}(1983)]{milgrom1983}
M.~{Milgrom}.
\newblock {A modification of the Newtonian dynamics as a possible alternative
  to the hidden mass hypothesis}.
\newblock \emph{Astrophysical Journal}, 270:\penalty0 365--370, 1983.
\newblock \doi{10.1086/161130}.

\bibitem[Morgan(1935)]{morgan1935}
T.~Morgan.
\newblock \emph{The relation of genetics to physiology and medicine}, pages
  1--16.
\newblock Stockholm: Imprimerie Royale, 1935.

\bibitem[Moss(2002)]{moss2002}
L.~Moss.
\newblock \emph{What Genes Can't Do}.
\newblock Cambridge, MA: The MIT Press, 2002.

\bibitem[Peebles(2020)]{peebles2020}
P.~Peebles.
\newblock \emph{Cosmology's Century: An Inside History of our Modern
  Understanding of the Universe}.
\newblock Princeton University Press, 2020.

\bibitem[Portin(1993)]{portin1993}
P.~Portin.
\newblock The concept of the gene: short history and present status.
\newblock \emph{Quart. Rev. Biol}, 68:\penalty0 173--223, 1993.

\bibitem[Quine(1951)]{quine1951}
W.~Quine.
\newblock Ontology and ideology.
\newblock \emph{Philosophical Studies}, 2:\penalty0 183--204, 1951.

\bibitem[Rheinberger et~al.(2015)Rheinberger, M\"{u}ller-Wille, and
  Meunier]{SEPgene}
H.-J. Rheinberger, S.~M\"{u}ller-Wille, and R.~Meunier.
\newblock {Gene}.
\newblock In E.~N. Zalta, editor, \emph{The {Stanford} Encyclopedia of
  Philosophy}. Metaphysics Research Lab, Stanford University, spring 2015
  edition, 2015.

\bibitem[Ruphy(2011)]{ruphy2011}
S.~Ruphy.
\newblock Limits to modeling: Balancing ambition and outcome in astrophysics
  and cosmology.
\newblock \emph{Simulation \& Gaming}, 42\penalty0 (2):\penalty0 177--194,
  2011.
\newblock \doi{10.1177/1046878108319640}.
\newblock URL \url{https://doi.org/10.1177/1046878108319640}.

\bibitem[Sandell(2010)]{sandell2010}
M.~Sandell.
\newblock Astronomy and experimentation.
\newblock \emph{Techn\'{e}}, 14\penalty0 (3):\penalty0 252--269, 2010.

\bibitem[Sanders and McGaugh(2002)]{sandersmcgaugh2002}
R.~H. Sanders and S.~S. McGaugh.
\newblock Modified newtonian dynamics as an alternative to dark matter.
\newblock \emph{Annual Review of Astronomy and Astrophysics}, 40\penalty0
  (1):\penalty0 263--317, 2002.
\newblock \doi{10.1146/annurev.astro.40.060401.093923}.
\newblock URL \url{https://doi.org/10.1146/annurev.astro.40.060401.093923}.

\bibitem[Scholz(2020)]{scholz2020}
E.~Scholz.
\newblock A scalar field inducing a non-metrical contribution to gravitational
  acceleration and a compatible add-on to light deflection.
\newblock \emph{General Relativity and Gravitation}, \penalty0 (46), 2020.

\bibitem[Shapere(1993)]{shapere1993}
D.~Shapere.
\newblock Discussion: Astronomy and antirealism.
\newblock \emph{Philosophy of Science}, 60:\penalty0 134--150, 1993.

\bibitem[Skordis and Z\l{}o\'{s}nik(2020)]{skordis2020}
C.~Skordis and T.~Z\l{}o\'{s}nik.
\newblock A new relativistic theory for modified {Newtonian} dynamics, 2020.

\bibitem[{Sterelny} and {Griffiths}(1999)]{sterelnygriffiths1999}
K.~{Sterelny} and P.~E. {Griffiths}.
\newblock \emph{Sex and Death: An Introduction to Philosophy of Biology}.
\newblock The University of Chicago Press, 1999.

\bibitem[Tabery(2019)]{SEPgenetics}
J.~Tabery.
\newblock {Genetics}.
\newblock In E.~N. Zalta, editor, \emph{The {Stanford} Encyclopedia of
  Philosophy}. Metaphysics Research Lab, Stanford University, fall 2019
  edition, 2019.

\bibitem[Turner(2018)]{turner2018}
M.~S. Turner.
\newblock {$\Lambda$CDM:} much more than we expected, but now less than what we
  want.
\newblock \emph{Foundations of Physics}, 48:\penalty0 1261--1278, 2018.
\newblock \doi{10.1007/s10701-018-0178-8}.

\bibitem[{Van Fraassen}(1980)]{vanfraassen1980}
B.~{Van Fraassen}.
\newblock \emph{The Scientific Image}.
\newblock Oxford: Oxford University Press, 1980.

\bibitem[Verlinde(2017)]{verlinde2017}
E.~Verlinde.
\newblock Emergent gravity and the dark universe.
\newblock \emph{SciPost Physics}, 2\penalty0 (3):\penalty0 016, 2017.

\bibitem[Zhao(2008)]{zhao2008}
H.~Zhao.
\newblock Reinterpreting {MOND:} coupling of {Einsteinian} gravity and spin of
  cosmic neutrinos?
\newblock 2008.

\end{thebibliography}

\end{document}